\documentclass[12pt]{article}

\usepackage{pb-diagram}
\usepackage{latexsym}
\usepackage{amsfonts}
\usepackage[usenames,dvipsnames]{xcolor}
\usepackage{graphicx,amssymb,amsmath,epsfig}
\usepackage[english]{babel}
\usepackage{graphicx}
\usepackage{dcolumn}
\usepackage{bm}

\definecolor{myblue}{rgb}{0,0,0.8}
\usepackage[colorlinks=true
,urlcolor=myblue	
,anchorcolor=myblue
,citecolor=myblue
,filecolor=myblue
,linkcolor=myblue
,menucolor=myblue
,pagecolor=myblue
,linktocpage=true  
]{hyperref}

\catcode`\@=11
\def\marginnote#1{}

\newcount\hour
\newcount\minute
\newtoks\amorpm
\hour=\time\divide\hour by60 \minute=\time{\multiply\hour by60
\global\advance\minute by-\hour}\edef\standardtime{{\ifnum\hour<12
\global\amorpm={am}%
        \else\global\amorpm={pm}\advance\hour by-12 \fi
        \ifnum\hour=0 \hour=12 \fi
        \number\hour:\ifnum\minute<10
0\fi\number\minute\the\amorpm}}
\edef\militarytime{\number\hour:\ifnum\minute<10 0\fi\number\minute}

\def\draftlabel#1{{\@bsphack\if@filesw {\let\thepage\relax
   \xdef\@gtempa{\write\@auxout{\string
      \newlabel{#1}{{\@currentlabel}{\thepage}}}}}\@gtempa
   \if@nobreak \ifvmode\nobreak\fi\fi\fi\@esphack}
        \gdef\@eqnlabel{#1}}
\def\@eqnlabel{}
\def\@vacuum{}
\def\draftmarginnote#1{\marginpar{\raggedright\scriptsize\tt#1}}
\def\draft{\oddsidemargin -.5truein
        \def\@oddfoot{\sl preliminary draft \hfil
        \rm\thepage\hfil\sl\today\quad\militarytime}
        \let\@evenfoot\@oddfoot \overfullrule 3pt
        \let\label=\draftlabel
        \let\marginnote=\draftmarginnote

\def\@eqnnum{(\theequation)\rlap{\kern\marginparsep\tt\@eqnlabel}%
\global\let\@eqnlabel\@vacuum}  }

\def\numberbysection{\@addtoreset{equation}{section}
        \def\theequation{\thesection.\arabic{equation}}}

\def\underline#1{\relax\ifmmode\@@underline#1\else
 $\@@underline{\hbox{#1}}$\relax\fi}

\catcode`@=12 \relax

\numberbysection

\topmargin by -\headsep
\def\nonu{\nonumber}
\def\br{\begin{eqnarray}}
\def\er{\end{eqnarray}}

\def\({\left(}
\def\){\right)}
\def\[{\left[}
\def\]{\right]}

\def\a{\alpha}

\def\b{\beta}

\def\d{\delta}

\def\bpsi{\bar{\psi}}

\def\g{\gamma}
\def\G{\Gamma}

\def\l{\lambda}
\def\L{\Lambda}

\def\o{\over}
\def\om{\omega}

\def\p{\phi}
\def\P{\Phi}
\def\pa{\partial}

\def\sech{\mathrm{sech}}

\def\tp0{\Theta_{+}^{(0)}}
\def\tm0{\Theta_{-}^{(0)}}

\def\bp{{\bar \p}}

\def\vep{\varepsilon}
\def\bvep{\bar{\varepsilon}}

\def\cK{{\cal K}}

\def\non{\nonumber}
\def\ba{\begin{align}}
\def\ea{\end{align}}
\def\be{\begin{eqnarray}}
\def\ee{\end{eqnarray}}
\def\L{\Lambda}
\def\a{\alpha}
\def\b{\beta}
\def\g{\gamma}
\def\d{\delta}
\def\P{\Psi}

\def\bp{\bar{\psi}}
\def\l{\lambda}
\def\G{\Gamma}
\def\ph{\phi}
\def\p{\psi}
\def\pp{\partial}
\def\o{\omega}

\begin{document}

\vspace*{1cm}
\noindent

\vskip 1 cm
\begin{center}
{\Large\bf  $N=1$ super sinh-Gordon model with defects revisited}
\end{center}
\normalsize
\vskip 1cm

\begin{center}
{A.R. Aguirre}\footnote{\href{mailto:aleroagu@ift.unesp.br}{aleroagu@ift.unesp.br}}, J.F. Gomes\footnote{\href{mailto:jfg@ift.unesp.br}{jfg@ift.unesp.br}},  N.I. Spano\footnote{\href{mailto:natyspano@ift.unesp.br}{natyspano@ift.unesp.br}} and A.H. Zimerman\footnote{\href{mailto:zimerman@ift.unesp.br}{zimerman@ift.unesp.br}}\\[1cm]

\par \vskip .1in \noindent
{Instituto de F\'isica Te\'orica - IFT/UNESP,\\
Rua Doutor Bento Teobaldo Ferraz, 271, Bloco II,
CEP 01140-070,\\ S\~ao Paulo - SP, Brasil.}\\
\vskip 3cm

\end{center}

\begin{abstract}

\noindent The Lax pair formalism is considered to discuss the integrability of the \mbox{$N=1$} supersymmetric sinh-Gordon model with a defect. We derive associated defect matrix for the model and construct the generating functions of the modified conserved quantities. The corresponding defect contributions for the modified energy and momentum of the model are explicitly computed.

\end{abstract}


\newpage
\tableofcontents

\vskip .4in

\section{Introduction}

Defects in integrable classical field theories have been an intensively studied topic in recent years \cite{Corr1}--\cite{FLZ2}. They were initially introduced in \cite{Corr1,Corr2} as internal boundary conditions described by a local Lagrangian density located at a fixed point, and it was shown for several type of bosonic field theories ,that these conditions correspond to frozen Backlund transformations and preserve integrability in the defect models. When the fields on either side of the defect only interact with each other at the boundary is referred to as a type-I defect.  However, it was shown in \cite{Corr09} that additional degrees of freedom associated to the defect itself can be also introduced through auxiliary fields which only exist at the defect point. These kind of defect are  named type-II defects.

More recently it was also suggested a different and rather systematic approach to defects in classical integrable field theories \cite{Cau}. The inverse scattering method formalism is used and the defect conditions are encoded in a defect matrix. This approach provided an elegant way to compute the modified conserved quantities, ensuring integrability. Using this framework the generating function for the modified conserved charges for any integrable evolution equation of the AKNS scheme were computed, and the type-II Backlund transformations for the sine-Gordon and Tzitz\'eica-Bullough-Dodd models had been also recovered \cite{Ale3}. The massive Thirring \cite{Ale4} and the supersymmetric Liouville \cite{Ale6} models have been also considered.

On the other hand, the question of involutivity of the modified conserved charges has been  addressed for several models \cite{Kundu}--\cite{Doi3}, by using essentially the algebraic framework of the classical $r$-matrix approach, and a modified transition matrix to describe integrable defects.

The presence of integrable defects in the $N=1$ supersymmetric sinh-Gordon (sshG) model  was discussed in \cite{FLZ1}, under the Lagrangian formalism and Backlund transformations. The authors have introduced a ``partially" type-II defect in the model since only a fermionic auxiliary field appears in the defect Lagrangian, and the integrability was considered in terms of zero curvature representation.

The purpose of this paper is to study the integrability of the $N=1$ sshG model from the defect matrix approach, in order to establish the existence of a generating function for an infinite set of modified conserved quantities. In section 2, we briefly review the defect $N = 1$ sshG
model, present the supersymmetry transformations that leave the total (bulk $+$ defect) action invariant and derive the defect contribution to the supercharge from the Lagrangian formalism. In section 3 we present the Lax pair formalism and derive explicitly the defect matrix for the $N=1$ sshG model. In section 4 we introduce the associated linear problem to discuss the conservation laws and then we compute the corresponding defect contributions to the modified conserved quantities. In section 5 we discuss further solutions for the defect matrix leading  in the special  case of the pure bosonic  limit to a  type-II Backlund transformation.  This fact suggests the derivation  of the  type-II Backlund transformation for the $N=1$ supersymmetric sinh-Gordon system.


\section{Review of the Lagrangian description}
The Lagrangian density describing the $N=1$ sshG model with type-I defects can be written as follows \cite{FLZ1},
\begin{equation}
 {\mathcal L} = \theta(-x)  {\mathcal L}_1 + \theta(x) {\mathcal L}_2 +\delta(x) \ {\mathcal L}_D, \label{e2.1}
\end{equation}
with
\begin{eqnarray}
 {\mathcal L}_p &=& \frac{1}{2}(\partial_x \phi_p)^2 - \frac{1}{2}(\partial_t \phi_p)^2 + \bar{\psi}_p(\partial_t -\partial_x)\bar{\psi}_p +   \psi_p(\partial_t +\partial_x)\psi_p  + V_p(\phi_p) \nonu \\ && + W_p(\phi_p,\psi_p, \bar{\psi}_p),\qquad\mbox{}\label{e2.2}\\[0.2cm]
   {\mathcal L}_D &=&\frac{1}{2}(\phi_2\partial_t\phi_1-\phi_1\partial_t\phi_2) -\psi_1\psi_2 -\bar{\psi}_1\bar{\psi}_2 + 2f_1\pa_t f_1 + B_0(\phi_1,\phi_2) \nonumber \\
   && + \,\,B_1(\phi_1,\phi_2,\psi_1,\psi_2,\bar{\psi}_1,\bar{\psi}_2,f_1),\quad \mbox{}\label{e2.3}
\end{eqnarray}
where  $\phi_p$ are real scalar fields, and $\psi_p,\bpsi_p$ are the components of  Majorana spinor fields in the regions $x<0$ ($p=1$) and $x>0$ ($p=2$) respectively, with the corresponding potentials given by
\begin{eqnarray}
 V_p&=& 4\cosh(2\phi_p),\qquad W_p = 8\bar{\psi}_p \psi_p \cosh\phi_p,\\
 B_0 &=& -\frac{4}{\om^2} \cosh(\phi_1+\phi_2) -\om^2\cosh(\phi_1-\phi_2) , \label{e2.5}\\
 B_1 &=& -\frac{4i}{\om}\cosh\left(\frac{\phi_1+\phi_2}{2}\right)f_1(\bar{\psi}_1+\bar{\psi}_2) +2i\om\,\cosh\left(\frac{\phi_1-\phi_2}{2}\right) f_1(\psi_1 -\psi_2).\qquad \mbox{}
\end{eqnarray}
The bulk fields equations are,
\begin{eqnarray}\label{mov1}
\partial_{x}^{2}\phi_{p}-\partial_{t}^{2}\phi_{p}&=&8\sinh(2\phi_{p})+8\,\bar{\psi}_{p}\psi_{p}\sinh\phi_{p},\nonumber\\
(\partial_{x}-\partial_{t})\bar{\psi}_{p}&=&4\,\psi_{p}\cosh\phi_{p},\nonumber\\
(\partial_{x}+\partial_{t})\psi_{p} &=&4\,\bar{\psi}_{p}\cosh\phi_{p}, \qquad p=1,2,
\end{eqnarray}
and the defect conditions at $x=0$ are given by
\begin{eqnarray}
\partial_{x}\phi_{1}-\partial_{t}\phi_{2} &=&\om^2\sinh(\phi_{1}-\phi_{2})+\frac{4}{\om^2}\sinh(\phi_{1}+\phi_{2})+\frac{2i}{\om}\,\sinh\left(\frac{\phi_{1}+\phi_{2}}{2}\right)f_{1}(\bar{\psi}_{1}+\bar{\psi}_{2})\nonumber\\
&&-i\om \,\sinh\left(\frac{\phi_{1}-\phi_{2}}{2}\right)f_{1}(\psi_{1}-\psi_{2}),\label{defeito1}\\
\partial_{x}\phi_{2}-\partial_{t}\phi_{1} & =& \om^2\sinh(\phi_{1}-\phi_{2})-\frac{4}{\om^2}\sinh(\phi_{1}+\phi_{2})-\frac{2i}{\omega}\,\sinh\left(\frac{\phi_{1}+\phi_{2}}{2}\right)f_{1}(\bar{\psi}_{1}+\bar{\psi}_{2})\nonumber\\
&& -i\om \,\sinh\left(\frac{\phi_{1}-\phi_{2}}{2}\right)f_{1}(\psi_{1}-\psi_{2}),
\label{defeito2}\\
\psi_{1}+\psi_{2} & = & -2i\om\,\cosh\left(\frac{\phi_{1}-\phi_{2}}{2}\right)f_{1},
\label{defeito3}\\
\bar{\psi}_{1}-\bar{\psi}_{2} & = & -\frac{4i}{\om}\,\cosh\left(\frac{\phi_{1}+\phi_{2}}{2}\right)f_{1},
\label{defeito4}\\
\partial_{t}f_{1} &=&\frac{i}{\om}\,\cosh\left(\frac{\phi_{1}+\phi_{2}}{2}\right)(\bar{\psi}_{1}+\bar{\psi}_{2})-\frac{i\om}{2}\,\cosh\left(\frac{\phi_{1}-\phi_{2}}{2}\right)(\psi_{1}-\psi_{2}).\quad \mbox{}
\label{defeito5} 
\end{eqnarray}
Here we have a fermionic degree of freedom $f_1$ at the defect which anticommutes with the fields $\psi_p$ and $\bar{\psi}_p$. If we also consider the $x$-derivative of $f_1$,
\begin{eqnarray}
\partial_{x}f_{1} &=&\frac{i}{\om}\,\cosh\left(\frac{\phi_{1}+\phi_{2}}{2}\right)(\bar{\psi}_{1}+\bar{\psi}_{2})-\frac{i\om}{2}\,\cosh\left(\frac{\phi_{1}-\phi_{2}}{2}\right)(\psi_{1}-\psi_{2}),\label{derxf1}
\end{eqnarray}
the eqns (\ref{defeito1})--(\ref{derxf1}) become the Backlund transformations for the supersymmetric sinh-Gordon model \cite{ChaiKul}. From this Lagrangian density we derive the canonical momentum
\begin{eqnarray}
 P=\int_{-\infty}^{0}\!\!dx\left(\pa_{t}\phi_{1}\pa_{x}\phi_{1}-\bpsi_{1}\pa_{x}\bpsi_{1}-\psi_{1}\pa_{x}\psi_{1}\right)+
\int_{0}^{+\infty}\!\!dx\left(\pa_{t}\phi_{2}\pa_{x}\phi_{2}-\bpsi_{2}\pa_{x}\bpsi_{2}-\psi_{2}\pa_{x}\psi_{2}\right),\quad
\end{eqnarray}
and by computing its time derivative, it was shown in \cite{FLZ1} that the modified momentum, namely,
\br
{\cal P} &=& P+ (\bpsi_{2}\bpsi_{1}+\psi_{1}\psi_{2})+\Big[\om^{2}\cosh(\phi_{1}-\phi_{2})-\frac{4}{\om^{2}}\cosh(\phi_{1}+\phi_{2})\Big]\nonu\\&&+\Big[\frac{4i}{\om}\cosh\left(\frac{\phi_{1}+\phi_{2}}{2}\right)(\bpsi_{1}+\bpsi_{2})+
2i\om\cosh\left(\frac{\phi_{1}-\phi_{2}}{2}\right)(\psi_{1}-\psi_{2})\Big]f_{1}\label{eqn2.15},
\er
is conserved after using properly the defect conditions  (\ref{defeito1})--(\ref{defeito5}). Analogously, for the energy,
\begin{eqnarray}
E&=&\int_{-\infty}^{0}\!\!dx\Big[\frac{1}{2}(\pa_{x}\phi_{1})^2+\frac{1}{2}(\pa_{t}\phi_{1})^2-\bpsi_{1}\pa_{x}\bpsi_{1}+\psi_{1}\pa_{x}\psi_{1}+4\cosh(2\phi_{1})+
8\bpsi_{1}\psi_{1}\cosh\phi_{1}\Big]\nonu \\&+&\!
\int_{0}^{+\infty}\!\!\! dx\Big[\frac{1}{2}(\pa_{x}\phi_{2})^2+\frac{1}{2}(\pa_{t}\phi_{2})^2-\bpsi_{2}\pa_{x}\bpsi_{2}+\psi_{2}\pa_{x}\psi_{2}+4\cosh(2\phi_{2})+ 8\bpsi_{2}\psi_{2}\cosh\phi_2\Big],\qquad\,\,\,
\end{eqnarray} 
the modified conserved energy is given by,
\br
\mathcal{E}&=&E+(\bpsi_{2}\bpsi_{1}-\psi_{1}\psi_{2})-\Big[\om^{2}\cosh(\phi_{1}-\phi_{2})+\frac{4}{\om^{2}}\cosh(\phi_{1}+\phi_{2})\Big]\nonu\\&& +\Big[
\frac{4i}{\om}\cosh\left(\frac{\phi_{1}+\phi_{2}}{2}\right)(\bpsi_{1}+\bpsi_{2})- 2i\om\cosh\left(\frac{\phi_{1}-\phi_{2}}{2}\right)(\psi_{1}-\psi_{2})\Big]f_{1}.\label{eqn2.17}
\er
In addition, the total action (bulk $+$ defect) for the defect sshG model is invariant under the supersymmetry (susy) transformations\footnote{{The light-cone coordinates are taken to be $x_{\pm}=x\pm t$, and therefore $\pa_{\pm}=\frac{1}{2}\left(\pa_x\pm \pa_t\right)$}},
\br
 \d_s\phi_p &=& \vep\,\psi_p+\bvep\,\bpsi_p, \label{3.1}\\
 \d_s\psi_p &=& -\vep \,\pa_- \phi_p + 2\bvep\sinh\phi_p,\\
 \d_s\bpsi_p &=& \bvep \,\pa_{+}\phi_p - 2\vep \sinh\phi_p , \qquad \quad p=1,2,\label{3.3}
\er
together with susy transformation of the auxiliary fermionic field \mbox{(see Appendix \ref{appA})},
\br 
\d_s f_1 &=& -i\omega \,\vep\sinh\(\frac{\phi_1-\phi_2}{2}\)  + \frac{2i \bvep}{\omega}\sinh\(\frac{\phi_1+\phi_2}{2}\).
\er
After introducing the defect at $x=0$, the corresponding  supercharges read as follows, 
\br
 Q_{\vep} &=&- \int_{-\infty}^{0}dx\,\left(2\psi_1\,\pa_-\phi_1 + 4\bpsi_1\sinh\phi_1\) -\int_{0}^{\infty}dx\,\left(2\psi_2\,\pa_-\phi_2 + 4\bpsi_2\sinh\phi_2\),\non\\
 \bar{Q}_{\bvep} &=&\int_{-\infty}^{0}dx\,\left(2\bpsi_1\,\pa_{+}\phi_1 + 4\psi_1\sinh\phi_1\)+\int_{0}^{\infty}dx\,\left(2\bpsi_2\,\pa_{+}\phi_2 + 4\psi_2\sinh\phi_2\). \mbox{}\label{e3.5}
\er
Now, by taking the time-derivative respectively, we get
\br
\frac{d Q_{\vep}}{dt} &=& \Big[2\psi_1\,\pa_-\phi_1 - 4\bpsi_1\sinh\phi_1-2\psi_2\,\pa_-\phi_2+ 4\bpsi_2\sinh\phi_2\Big]_{x=0}
\er
and 
\br
\frac{d \bar{Q}_{\vep}}{dt} &=& \Big[2\bpsi_1\,\pa_{+}\phi_1 -4\psi_1\sinh\phi_1-2\bpsi_2\,\pa_{+}\phi_2 +4\psi_2\sinh\phi_2\Big]_{x=0}.
\er
Then, by considering the defect conditions (\ref{defeito1})-- (\ref{defeito5}), we find that the modified conserved supercharges take the following
 form \cite{Ale5},
\br
 {\cal Q} &=& Q_\vep + Q_{\mbox{\tiny D}}, \qquad \mbox{and} \qquad \bar{\cal Q} = \bar{Q}_{\bvep} + \bar{Q}_{\mbox{\tiny D}},
\er
where
 \br
 Q_{\mbox{\tiny D}} &=& -4i\omega\left[\sinh\Big(\frac{\phi_1-\phi_2}{2}\Big)\,f_1 \right]_{x=0}, \quad \mbox{and} \quad  \bar{Q}_{\mbox{\tiny D}}= \frac{8i}{\om} \left[\sinh\Big(\frac{\phi_1+\phi_2}{2}\Big)\,f_1 \right]_{x=0}.\qquad \mbox{}\label{e3.7}
\er
%

\section{Lax formulation and defect matrix}

The $N=1$ sshG equation can be derived as a compatibility condition of the following linear system of first-order differential equations, 
\br\label{problem}
\pa_{\pm}\Psi (x_\pm,\l) =-\mathcal{A}_{\pm}\Psi(x_\pm,\l) ,
\er
where $\Psi(x_\pm,\l) $ is a three-component vector-valued field, $\l$ is the spectral parameter, and the Lax connections ${\cal A}_\pm$ are $3 \times 3 $ graded matrices valued in the $sl(2,1)$ Lie superalgebra, which can be written in the following form,
\br
\mathcal{A}_{+} &=&\left(\begin{array}{cc|c}\lambda^{1/2}-\pa_{+}\phi&-1&\bar{\psi}\\[0.2cm] -\l&\l^{1/2}+\pa_{+}\phi&\l^{1/2}\bpsi \tabularnewline\hline \mbox{}&\mbox{}&\mbox{}\\[-0.3cm]
 \l^{1/2}\bpsi &\bpsi &2\l^{1/2}\end{array}\right)\label{lax +}, \\[0.2cm]
\mathcal{A}_{-}&=&\left(\begin{array}{cc|c}\lambda^{-1/2} &-\lambda^{-1}e^{2\phi}&-\lambda^{-1/2}\psi \,e^{\phi}\\[0.2cm]- e^{-2\phi}&\lambda^{-1/2}&- \psi e^{-\phi} \tabularnewline\hline \mbox{}&\mbox{}&\mbox{}\\[-0.3cm]
\psi e^{-\phi}&\lambda^{-1/2}\psi \,e^{\phi}&2\lambda^{-1/2}\end{array}\right)\label{lax -}.
\er
Then, from the zero-curvature condition or Zakharov-Shabat equation,
\br
\pa_{+}\mathcal{A}_{-}-\pa_{-}\mathcal{A}_{+}+\left[\mathcal{A}_{+},\mathcal{A}_{-}\right]=\ 0,
\er 
we  recover the sshG field  equations (\ref{mov1}). Now, to derive the defect matrix via gauge transformations, we consider the existence  of a graded matrix ${\cal K}$ connecting two different configu\-rations, namely $\Psi^{(2)}= {\cK}(\l)\Psi^{(1)}$, satisfying the following equations,
\br\label{gauge}
\pa_{\pm}\cK=\cK\mathcal{A}^{(1)}_{\pm}-\mathcal{A}^{(2)}_{\pm}\cK,
\er
where ${\cal A}_{\pm}^{(p)}$ represents the Lax connections depending on the respective fields $\phi_p, \psi_p$, and $\bpsi_p$. Let us consider the following ansatz for the $\l$-expansion of the matrix $\cK$,
\begin{eqnarray}\label{K}
\cK_{ij}=\a_{ij}+\l^{-1/2}\b_{ij}+\l^{1/2}\g_{ij},
\end{eqnarray}
with $\a_{ij}, \b_{ij}$, and $\g_{ij}$ being the entries of $3 \times 3$ graded matrices. First of all, by considering the $\l$-expansion in order to solve the differentials equations (\ref{gauge}), we find that the $\l^{+ 3/2}$ and $\l^{+ 1}$  terms  lead to
\begin{eqnarray}
\a_{12}=\g_{12}=\g_{13}=\g_{32}=0, \qquad \quad \g_{11}=\g_{22}=c_{11}, \qquad \g_{33}=c_{33}, 
\end{eqnarray}
and 
\begin{eqnarray}
 \a_{11}-\a_{22}&=&\bp_2\,\g_{31}-\g_{23}\,\bp_1,\label{vin1}\\
\a_{13}+\g_{23}&=&\bp_2\,\g_{33}-\g_{11}\,\bp_1,\label{vin2}\\ 
\g_{31}+\a_{32}&=&\bp_1\,\g_{33}-\g_{11}\,\bp_2,\label{vin3}
\er
where $c_{ij}$ denotes arbitrary constants\footnote{In what follows we will denote all the constants with Latin letters}. Analogously, for the degrees  $\l^{-3/2}$ and $\l^{-1}$ we get 
\begin{eqnarray}
\a_{21}=\b_{21}=\b_{23}= \b_{31}=0, \qquad \quad \b_{22}=\b_{11}e^{2\phi_-}, \qquad \b_{33}=b_{33},\label{eqn3.11}
\end{eqnarray}
and the following constraints,
\br
\a_{11}\,e^{\phi_-}-\a_{22}\,e^{-\phi_-}&=&e^{-\frac{\phi_+}{2}}\big(\b_{13}\,\p_1\,e^{\frac{\phi_-}{2}}+e^{-\frac{\phi_-}{2}}\,\p_2\,\b_{32}\big),\label{eqn 3.12}\\
\a_{31}\,e^{{(\phi_+ +\phi_-)}}+\b_{32}&=& e^{\frac{\phi_+ }{2}}\big(\b_{33}\,\p_1\,e^{\frac{\phi_-}{2}}-\b_{22}\,\p_2\,e^{-\frac{\phi_-}{2}}\big),\label{eqn 3.13}\\
\a_{23}\,e^{(\phi_+-\phi_-)}+ \b_{13}&=&e^{\frac{\phi_+}{2}}(\b_{11}\,\p_1e^{\frac{\phi_-}{2}}-\b_{33}\,\p_2 e^{-\frac{\phi_-}{2}}).\label{eqn 3.14}
\er
We have denoted $\phi_{\pm}=\phi_1\pm \phi_2$. Notice that after suitable parametrizations the constraints (\ref{vin2}), (\ref{vin3}), (\ref{eqn 3.13}) and (\ref{eqn 3.14}) could reproduce the fermionic defect conditions (\ref{defeito3}) and (\ref{defeito4}) respectively, by introducing properly the auxiliary field $f_1$. The bosonic defect conditions will be derived from the differential equations coming from the degrees $\l^{0}$ and $\l^{\pm 1/2}$, which  are fully presented in appendix \ref{appB}. Now, after considering the equations involving $\b_{11}$ (\ref{dm39}) and $\b_{22}$ (\ref{dm42}), namely,
\br
 \pa_+\b_{11} = -\b_{11} \pa_+\phi_-, \qquad \pa_+\b_{22} = \b_{22} \pa_+\phi_-, 
\er
and the constraint in (\ref{eqn3.11}), we find the simple solution $\b_{11} = b_{11} \,e^{-\phi_-}$ and $\b_{22}= b_{11}\,e^{\phi_-}$. Now, by setting $\b_{33}= b_{33}=-b_{11}$ and  $c_{33}=c_{11}$, we get from (\ref{vin2}) and (\ref{vin3}) that
\br
 \a_{13} +\a_{32}= -(\g_{23}+\g_{31}).\label{rel 3.16}
\er
From the eqs (\ref{dm27}), (\ref{dm31}), (\ref{dm47}) and (\ref{dm48}), involving both sides of the relation ({\ref{rel 3.16}), we find that
\br
 \pa_-(\a_{13} +\a_{32}) =0, \qquad \pa_-(\g_{23}+\g_{31})=0.
\er
Then, we will consider the simple solution when the components satisfy that $\a_{32} =-\a_{13}$, and  $\g_{23}=-\g_{31}$. Using
\br
\b_{11} = b_{11} e^{-\phi_-}, \quad {\rm {and}}\quad \b_{33} = -b_{11},
\er 
eqn. (\ref{eqn 3.14}) gives
\be
\b_{13}e^{-(\phi_+-\phi_-)}+\a_{23}&=&b_{11}e^{-\frac{(\phi_+-\phi_-)}{2}}(\p_1+\p_2), \label{3.18}
\ee
Introducing the auxiliary field $f_1$,
\begin{eqnarray}
f_{1}=\frac{i}{2\o}{\sech\Big(\frac{\phi_-}{2}\Big)}\p_{+}
=\frac{i\o}{4}\sech\Big(\frac{\phi_+}{2}\Big)\bp_{-},\label{eqn3.19}
\end{eqnarray}
we find that eqn. (\ref{3.18})  becomes
\be
\a_{23}e^{\frac{(\phi_+-\phi_-)}{2}}+\b_{13}e^{-\frac{(\phi_+-\phi_-)}{2}}&=&-2i\o b_{11}\cosh\left(\frac{\phi_-}{2}\right)f_{1}.
\ee
After making the choice 
\br \b_{13}=\a_{23}e^{\phi_+} \label{beta13}\er
 with 
\be
\a_{23}=-i\o b_{11}e^{-\frac{\phi_+}{2}}f_{1}.
\ee
In the same way, by taking eq (\ref{vin1})--(\ref{eqn 3.13}) we obtain 
\br
\a_{13}=\frac{2i}{\omega}c_{11}\,e^{\frac{\phi_+}{2}}f_1, \quad \a_{31}=i\o b_{11}e^{-\frac{\phi_+}{2}}f_1,\quad \b_{32}=\a_{31}\, e^{\phi_+},\quad\g_{23}=\a_{13}\,e^{-\phi_+},\label{3.22}
\er
and
\br
\a_{11} = \a_{22}= \frac{i\omega b_{11}}{4}\sech\Big(\frac{\phi_-}{2}\Big)\psi_-\,f_1.\label{eqn 3.23}
\er
From the eqs (\ref{dm19}) and (\ref{dm22}), we also have 
\be
2(\pp_{+}\a_{11})&=&(\b_{13}+\a_{23})\bp_1-\bp_2(\a_{31}+\b_{32})\non\\
&=&-\frac{\o^2b_{11}}{2} \left(\bp_2\bp_1-\bp_2\bp_1\right)  \non\\
&=&0,
\ee
where we have used for $\b_{13}$ and $\b_{32}$ eqns. (\ref{beta13}) and (\ref{3.22}) respectively.

Similarly, from (\ref{dm26}) and (\ref{dm28}) we obtain  $\pp_{-}\a_{11}=\pp_{-}\a_{33}=\pp_{+}\a_{33}=0$. Therefore, it is suitable to set $\a_{11}=a_{11}$ and $\a_{33}=a_{33}$.
Now, let us consider the eqs (\ref{dm33}), (\ref{dm36}), (\ref{dm50}), and (\ref{dm51}) involving derivatives of the elements $\b_{11}, \b_{22}, \g_{11}$, and $\g_{22}$, 
\begin{eqnarray}
\pp_{+}\phi_- &=& \frac{1}{c_{11}}\left[\g_{21}-\b_{12}-c_{11}\bp_2\bp_{1}+\a_{13}\bp_{1}-\bp_{2}\a_{13}\right]\non\\&=&\frac{1}{c_{11}}(\g_{21}-\b_{12}) -\frac{2i}{\omega} \sinh\Big(\frac{\phi_+}{2}\Big)  \bpsi_+\,f_1, \label{eqn3.21}\\
\pp_{-}\phi_- &=&\frac{1}{b_{11}}\left[\b_{12}e^{-\phi_+}-\g_{21}e^{\phi_+}-b_{11}\p_2\p_1+\a_{23}\p_1e^{\frac{(\phi_+-\phi_-)}{2}}-\p_2\a_{31}e^{\frac{(\phi_+-\phi_-)}{2}}\right]\non \\&=&\frac{1}{b_{11}}(\b_{12}e^{-\phi_+} -\g_{21}e^{\phi_+}),
\end{eqnarray}
and eqs (\ref{dm34}), (\ref{dm40}), (\ref{dm46}) and (\ref{dm52}) for $\b_{12}$ and $\g_{21}$ respectively,
\begin{eqnarray}
\pp_{+}\b_{12}&=&\b_{12}(\pp_{+}\phi_+)+2b_{11}\sinh \phi_{-}, \\[0.1cm]
\pp_{+}\g_{21}&=&-\g_{21}(\pp_{+}\phi_+)-2b_{11}\sinh \phi_{-},\\[0.1cm]
\pp_{-}\b_{12}&=&-2c_{11}e^{\phi_+}\sinh\phi_- -\frac{2ic_{11}}{\omega} \,e^{\phi_+}\sinh\Big(\frac{\phi_-}{2}\Big)\psi_- \,f_1, \label{eqn3.25}\\[0.1cm]
\pp_{-}\g_{21}&=&2c_{11}e^{-\phi_+}\sinh\phi_- +\frac{2ic_{11}}{\omega} \,e^{-\phi_+}\sinh\Big(\frac{\phi_-}{2}\Big)\psi_- \,f_1. \label{eqn3.26}
\end{eqnarray}
By introducing the following parametrizations,
\br
 \b_{12} = b_{12} \,e^{\phi_+}, \qquad \g_{21} = b_{12} \,e^{-\phi_+}, \qquad b_{12} = -\frac{2c_{11}}{\omega^2}, \qquad b_{11}=0,
\er
the eqs (\ref{eqn3.21}), (\ref{eqn3.25}) and (\ref{eqn3.26}) become exactly the defect conditions for the bosonic field $\phi_{\pm}$, namely,
\br
\pp_{+}\phi_-&=&\frac{4}{\omega^2} \sinh\phi_+ -\frac{2i}{\omega} \sinh\Big(\frac{\phi_+}{2}\Big)  \bpsi_+\,f_1, \\
\pp_{-}\phi_+&=&\omega^2\sinh\phi_- +i{\omega} \sinh\Big(\frac{\phi_-}{2}\Big)\psi_- \,f_1. \label{eqn3.33}
\er
Since $b_{11}=0$, we also find that the set of elements $\{\a_{11},\a_{22},\a_{23},\a_{31}, \b_{11},\b_{13}, \b_{22},\b_{32},\b_{33}\}$ completely vanish and then do not contribute at all to the ${\cal K}$ matrix. 

Finally, if we consider equation (\ref{dm30}) involving the element $\a_{33}$, we obtain
\br
\pp_{-}\a_{31}&=&-\g_{31}-\a_{32}e^{-(\phi_+ + \phi_-)}+\a_{33}\p_1\,e^{-\frac{(\phi_++\phi_-)}{2}}-\p_2\big(e^{-\frac{(\phi_+-\phi_-)}{2}}\a_{11}+e^{\frac{(\phi_+-\phi_-)}{2}}\g_{21}\big),\non\\
0&=&e^{-\frac{\phi_+}{2}}\left[\frac{2i}{\omega}c_{11}\big (1+e^{-{\phi_-}}\big)f_1+e^{-\frac{\phi_-}{2}}\big(\a_{33}\psi_1 -b_{12}\psi_2\big)\right],
\er
from where we conclude that $\a_{33} ={2c_{11}}{\omega^{-2}}$.

Therefore, we have found a suitable solution for the defect matrix ${\cal K}$, which can be written in the following form,
\begin{eqnarray}\label{K caso bf}
\cK=\left(\begin{array}{cc|c}
  c_{11} \,\l^{1/2}& -\frac{2}{\omega^2}\,c_{11}\, e^{\phi_+}\,\lambda^{-1/2} & \frac{2i}{\om} c_{11} e^{\frac{\phi_+}{2}}f_1\\[0.3cm]
 -\frac{2}{\omega^2}\,c_{11}\,e^{-\phi_+} \lambda^{1/2} & c_{11} \,\lambda^{1/2}& \frac{2i}{\om}c_{11}e^{-\frac{\phi_+}{2}}f_1 \, \lambda^{1/2}\tabularnewline\hline \mbox{}&\mbox{}&\mbox{}\\[-0.2cm]
-\frac{2i}{\om}\,c_{11}\,e^{-\frac{\phi_+}{2}}f_1\,\lambda^{1/2} & -\frac{2i}{\om}c_{11}e^{\frac{\phi_+}{2}}f_1 & \frac{2}{\omega^2}\,c_{11}+c_{11}\,\lambda^{1/2} \\
\end{array} \right),
\end{eqnarray}
where $\omega$ represent the Backlund parameter and $c_{11}$ is a free constant parameter. Now, as it was proposed in \cite{Ale4}, the defect matrix will be used to derive modified conserved quantities.
To do that, in the next section we will derive the bulk energy and momentum by using the Lax formalism.\\[-0.4cm]


\section{Conservation laws}

In this section we will construct explicitly  generating functions for an infinite set of independent conserved quantities for the sshG model in the bulk theory, as well as derive the corresponding modified conserved quantities arising from the defect contributions in the defect theory, by using the Lax approach.

\subsection{Associated linear problem and conserved quantities}

Let us consider the associated linear problem for the sshG model in the $(x,t)$ coordinates as follows,
\begin{eqnarray}
\pp_{t}\P(x,t;\l)&=&U(x,t;\l)\P(x,t;\l), \label{problema}\\
\pp_{x}\P(x,t;\l)&=&V(x,t;\l)\P(x,t;\l),\label{problema2}
\end{eqnarray}
with $U=\mathcal{A}_{-}-\mathcal{A}_{+}$ and $V=-(\mathcal{A}_{+}+\mathcal{A}_{-})$ are respectively determined as linear combinations of the Lax connections (\ref{lax +}) and (\ref{lax -}), and the vector-valued function $\Psi$ has the form $\Psi=(\Psi_1, \Psi_2, \epsilon\Psi_3)^T$, with bosonic components $\Psi_i$ and $\epsilon$ a Grassmannian parameter. 

Now, as it was claimed in \cite{Wadati}, it is possible to construct a generating function for the  conservation laws by defining a set of auxiliary functions $\G_{ij}[\P]=\P_i\P_j^{-1}$, for $i,j=1,2,3$. Then, considering the linear system (\ref{problema}) and (\ref{problema2}), we find that the $j$-th conservation equation can be written as,
\br
 \pa_t\Bigg[V_{jj} + \sum_{i\neq j} V_{ji} \,\G_{ij}\Bigg] = \pa_x\left[U_{jj} + \sum_{i\neq j} U_{ji}\, \G_{ij}\],\label{law}
\er
where the functions $\G_{ij}$ satisfy the following Riccati equations,
\br
 \pa_x \G_{ij} &=& \left(V_{ij}-V_{jj}\G_{ij}\) + \sum_{k\neq j}\big(V_{ik} - \G_{ij} \,V_{jk}\big) \G_{kj} \,,\label{e4.4}\\
  \pa_t \G_{ij} &=& \left(U_{ij}-U_{jj}\G_{ij}\) + \sum_{k\neq j}\big(U_{ik} - \G_{ij}U_{jk}\big)\G_{kj}.\label{e4.5}
\er
Then, the corresponding $j$-th generating function of the conserved quantities reads, 
\br
 { I}_j = \int_{-\infty}^{\infty}dx\,\left[V_{jj} + \sum_{i\neq j} V_{ji} \,\G_{ij}\]. \label{e4.6}
\er
In order to derive explicitly the conserved quantities, it is necessary to introduce  $\l$-expansions of   the functions $\G_{ij}$ in positive and negative powers of the spectral parameter to solve in a recursive way the Riccati equations for each coefficient. As a consequence, an infinite set of conserved quantities will appear from the generating function (\ref{e4.6}). In particular, to derive the energy and momentum we will consider the $\l^{1/2}$-terms of the charges $I_1$ and $I_2$.

\noindent Firstly, let us consider the explictly form of the Riccati equations for $j=1$,
\begin{eqnarray}\label{ricatti linha 1}
\pp_{x}\G_{21}&=&\l+e^{-2\ph}-(\pp_{x}\phi+\pp_{t}\phi)\ \G_{21}-
\left(\l^{1/2}\bp-e^{-\ph}\p\right)\G_{31}-
\left(1+\l^{-1}e^{2\ph}\right)\left(\G_{21}\right)^2\non\\&&+
\left(\bp-\l^{-1/2} e^{\ph}\p\right) \G_{21}\G_{31},\label{ricatti linha 1}\\
\pp_{x}\G_{31}&=&-\l^{1/2}\bp-\p e^{-\ph}-\left(\frac{1}{2}(\pp_{x}\phi+\pp_{t}\phi)+\l^{1/2}+\l^{-1/2}\right)\G_{31}-\left(\bp+\l^{-1/2}\p e^{\ph}\right) \G_{21}\non\\&&-
\left(1+\l^{-1}e^{2\ph}\right) \G_{31}\G_{21} .\label{ricatti linha 1,1}
\end{eqnarray}
Now, we expand $\G_{12}$ and $\G_{31}$ as $\l\to 0$,
\begin{eqnarray}
\G_{21}=\sum_{n=0}^{\infty}\l^{n/2}\G_{21}^{(n/2)},\qquad\G_{31}=\sum_{n=0}^{\infty}\l^{n/2}\G_{31}^{(n/2)} .
\end{eqnarray}
By inserting these expansions into the Riccati equations (\ref{ricatti linha 1}) and (\ref{ricatti linha 1,1}) we find that the first coefficients are given by,
\br
 \G_{21}^{(0)}&=&0, \qquad \G_{31}^{(0)}=0, \qquad \G_{21}^{(1/2)}= e^{-2\ph},\qquad \G_{31}^{(1/2)}=-e^{-\ph}\p,\label{eqn4.10}\\[0.1cm]
\G_{21}^{(1)}&=&\frac{1}{2}\ e^{-2\ph}(\pp_{x}\phi-\pp_{t}\phi),\\[0.1cm]
\G_{31}^{(1)}&=&\frac{1}{2}\ e^{-\ph}\left[\pp_{x}\p -2\bp\cosh\phi-\frac{1}{2}\psi(\pp_{x}\ph-\pp_{t}\ph)\right],\\[0.1cm]
\G_{21}^{(3/2)}&=& e^{-2\ph}\left[\frac{1}{8}(\pp_{x}\phi-\pp_{t}\phi)^2-
\frac{1}{4}\pp_{x}(\pp_{x}\phi-\pp_{t}\phi)+\sinh(2\phi)+\bp\p\sinh\phi\].\label{eqn4.13}
\er
Thus, we find from the first generating function of conserved quantities (\ref{e4.6}), namely
\br
I_{1}&=& \int _{-\infty}^{\infty}\!\!dx \Big[-(\l^{1/2}+\l^{-1/2})+\frac{1}{2}(\pp_{x}\phi+\pp_{t}\phi)+\left(1+\l^{-1}e^{2\ph}\right)\G_{21}\non\\&&\qquad \quad\,\, +
\left(\l^{-1/2}e^{\ph}\p-\bp\right)\G_{31}\Big],\quad \mbox{} \label{e4.14}
\er
that the charge $I_1^{(-1/2)}$ vanishes, $I_1^{(0)}$ is the topological charge, namely,
\br
 I_1^{(0)} = \int _{-\infty}^{\infty} dx \ \pa_x\phi, \label{eqn4.15}
\er
and the first non-trivial conserved quantity is given by the $\l^{1/2}$-term, which is given by
\begin{eqnarray}\label{I+1}
I_{1}^{(1/2)}&=& \int _{-\infty}^{\infty}dx\ \left[\frac{1}{8}(\pp_{x}\ph-\pp_{t}\ph)^2+\frac{1}{2}\p\,\pp_{x}\p+2\bp\p\cosh\ph +(\cosh 2\ph-1)\right].
\end{eqnarray}
Now, if we consider the expansion of the $\G_{12}$ and $\G_{31}$ as $\l \to \infty$, 
\begin{eqnarray}
\G_{21}=\sum_{n=-1}^{\infty}\frac{\hat{\G}_{21}^{(n/2)}}{\l^{n/2}},\qquad\G_{31}=\sum_{n=0}^{\infty}\frac{\hat{\G}_{31}^{(n/2)}}{\l^{n/2}},
\end{eqnarray}
we get,
\begin{eqnarray}
\hat{\G}_{31}^{(0)}&=&-\bp, \qquad \hat{\G}_{21}^{(0)}= -\frac{1}{2}(\pp_{x}\phi+\pp_{t}\phi),\qquad \hat{\G}_{21}^{(-1/2)}= 1,\label{eqn4.17}\\
\hat{\G}_{21}^{(1/2)}&=&\frac{1}{8}(\pp_{x}\phi +\pp_{t}\phi)^2+\frac{1}{4}\pa_x(\pp_{x}\phi+\pp_{t}\phi) -\sinh(2\ph) +\p\bp\sinh\ph,\\
\hat{\G}_{31}^{(1/2)}&=&\frac{1}{2}\left[\pp_{x}\bp -2\p\cosh\ph + \frac{1}{2}\,\bp\,(\pp_{x}\phi +\pp_{t}\ph)\right].\label{eqn4.19}
\end{eqnarray}
Then, by introducing these results into eq. (\ref{e4.14}) we find in this case that the zero-order term vanishes and the first non-vanishing conserved quantity is given by the $\l^{-1/2}$ order, as follows
\begin{eqnarray}\label{I-1}
\hat{I}_{1}^{(-1/2)}&=&\int _{-\infty}^{\infty}dx\ \left[
\frac{1}{8}\left(\pp_{x}\phi+\pp_{t}\phi\right)^2-\frac{1}{2}\bp\,\pp_{x}\bp+2\bp\p\cosh\ph+
\left(\cosh 2\phi-1\)\right].\quad \mbox{}
\end{eqnarray}
Let us now consider the respective Riccati equations for $j=2$, 
\begin{eqnarray}
\pp_{x}\G_{12}&=& 1+\l^{-1}e^{2\ph}+(\pp_{x}\phi+\pp_{t}\phi)\ \G_{12}+\left(\l^{-1/2}e^{\ph}\p-\bp\right)\G_{32}-\left(\l+e^{-2\ph}\right)\left(\G_{12}\right)^2\non\\&&
+\left(\l^{1/2}\bp- e^{-\ph}\p\right)\G_{12}\G_{32}\label{ricatti 2},\\
\pp_{x}\G_{32}&=&-\bp-\l^{-1/2} e^{\ph}\p+\left(\frac{1}{2}(\pp_{x}\phi+\pp_{t}\phi)-\l^{1/2}-\l^{-1/2}\right)\G_{32}-\left(\l^{1/2}\bp+\p e^{-\ph}\right)\G_{12}\non\\&&
-\left(\l+e^{-2\ph}\right)\G_{32}\G_{12}.\label{ricatti 2,2}
\end{eqnarray}
As it was done before, we expand $\G_{12}$ and $\G_{32}$ as $\l\to0$, 
\begin{eqnarray}
\G_{12}=\sum_{n=-1}^{\infty}\l^{n/2}\G_{12}^{(n/2)},\qquad
\G_{32}=\sum_{n=0}^{\infty}\l^{n/2}\G_{32}^{(n/2)},
\end{eqnarray}
and then we get
\begin{eqnarray}
\G_{12}^{(0)}&=& -\frac{1}{2}\ e^{2\ph}\left(\pp_{x}\phi-\pp_{t}\phi\right), \qquad \G_{32}^{(0)}= -e^{\ph}\p, \qquad \G_{12}^{(-1/2)}=e^{2\ph}, \label{eqn4.24}\\
\G_{12}^{(1/2)}&=&e^{2\ph}\left[\frac{1}{8}\left(\pp_{x}\phi-\pp_{t}\phi\right)^2+
\frac{1}{4}\pa_x\left(\pp_{x}\phi-\pp_{t}\phi\right) -\sinh(2\phi)-\bp\p\sinh\phi\right],\qquad \mbox{}\\
\G_{32}^{(1/2)}&=&\frac{1}{2}e^{\ph}\left[\pp_{x}\p-2\bp\cosh\phi+\frac{1}{2}\psi (\pp_{x}\phi-\pp_{t}\phi)\right].\label{eqn4.26}
\end{eqnarray}
From the generating function (\ref{e4.6}) for $j=2$,
\br
I_{2}=\int _{-\infty}^{\infty}dx \Big[-\l^{1/2}-\l^{-1/2}-\frac{1}{2}(\pp_{x}\phi+\pp_{t}\phi)+\left(\l+e^{-2\ph}\right)\G_{12}+\left(e^{-\ph}\p-\l^{1/2}\bp\right)\G_{32}\Big],\quad \mbox{}\label{e4.28}
\er
we find again that the zero-order gives us the topological term but in this case with an opposite sign,
\br
 I_2^{(0)} = -\int _{-\infty}^{\infty}dx \ \pa_x\phi,\label{eqn4.29}
\er
and the first non-trivial conserved quantity is given by the $\l^{1/2}$-order in the following way,
\br\label{I+2}
I_{2}^{(1/2)}&=&\int _{-\infty}^{\infty}dx \Big[\frac{1}{8}\left(\pp_{x}\phi-\pp_{t}\phi\)^2+\frac{1}{2}\p\,\pp_{x}\p+2\bp\p\cosh\ph+\left(\cosh 2\ph-1\)\Big].
\er
Correspondingly, for the expansion around $\l\to \infty$,
\begin{eqnarray}
\G_{12}=\sum_{n=0}^{\infty}\frac{\hat{\G}_{12}^{(n/2)}}{\l^{n/2}},\qquad
\G_{32}=\sum_{n=0}^{\infty}\frac{\hat{\G}_{32}^{(n/2)}}{\l^{n/2}},
\end{eqnarray}
we find,
\begin{eqnarray}
\hat{\G}_{32}^{(0)}&=&0,\qquad \hat{\G}_{12}^{(1/2)}=1, \qquad \hat{\G}_{32}^{(1/2)}=-\bp, \label{eqn4.30}\\
\hat{\G}_{12}^{(1)}&=&\frac{1}{2}(\pp_{x}\phi+\pp_{t}\phi), \\
\hat{\G}_{12}^{(3/2)}&=&\frac{1}{8}\left(\pp_{x}\phi+\pp_{t}\phi\)^2
-\frac{1}{4}(\pp_{x}^2+\pp_{x}\pp_{t})\ph + \sinh(2\ph)+\bp\p\sinh\phi ,\\
\hat{\G}_{32}^{(1)}&=&\frac{1}{2}\left[\pp_{x}\bp -2\p\cosh\ph-\frac{1}{2}\bp(\pp_{x}\phi+\pp_{t}\phi)\right] .\label{eqn4.33}
\end{eqnarray}
From eq (\ref{e4.28}) we obtain that the zero-order term vanishes and the first non-vanishing conserved quantity is given by,
\begin{eqnarray}\label{I-2}
\hat{I}_{2}^{(-1/2)}&=&\int _{-\infty}^{\infty}dx\ \Big[\frac{1}{8}\left(\pp_{x}\phi+\pp_{t}\phi\)^2-\frac{1}{2}\bp\,\pp_{x}\bp +2\bp\p\cosh\ph +\left(\cosh 2\ph-1\)\Big].
\end{eqnarray}
Notice that by adding (\ref{eqn4.15}) and (\ref{eqn4.29}) we find that the topological charge of the model is zero \cite{VF77}, \cite{Hruby}. On the other hand $I_{2}^{(1/2)}=I_{1}^{(1/2)}$ and $\hat{I}_{2}^{(-1/2)}=\hat{I}_{1}^{(-1/2)}$, and then we can introduce a new set of conserved quantities defined by
\be
\mathbb{I}^{(+1/2)}=I^{(1/2)}_{1}+I^{(1/2)}_{2},\quad
\hat{\mathbb{I}}^{(-1/2)}=\hat{I}^{(-1/2)}_{1}+ \hat{I}^{(-1/2)}_{2} .
\ee
Then, the energy and momentum for the sshG model in the bulk theory are  recovered in this formalism through the following combinations of the conserved quantities defined above,
\begin{eqnarray}
E=\hat{\mathbb{I}}^{(-1/2)}+\mathbb{I}^{(+1/2)}, \qquad P =\hat{\mathbb{I}}^{(-1/2)}-\mathbb{I}^{(+1/2)},
\end{eqnarray}
with,
\br
E&=& \int_{-\infty}^{\infty}dx\,\Big[\frac{1}{2}(\pp_{x}\ph)^2+\frac{1}{2}(\pp_{t}\ph)^2-\bp\,\pp_{x}\bp+\p\,\pp_{x}\p+4\left(\cosh 2\ph-1\) +8\bp\p\cosh\ph\Big],\qquad\mbox{}\\
P&=& \int_{-\infty}^{\infty}dx\ \left[\pp_{x}\ph\,\pp_{t}\ph-\p\,\pp_{x}\p-\bp\,\pp_{x}\bp\right].
\er
In the next subsection we will derive explicitly the corresponding defect contribution to the modified energy and momentum after introducing the defect in the formalism.


\subsection{Defect contributions}

So far we have considered the derivation of conserved quantities by using the Lax formalism in the bulk theory. In this section we will consider the modification to those quantities after introducing the defect into the formalism. To do that, let us recall how to construct the corresponding modified conserved quantities from the defect matrix. 

As explained in \cite{Ale4}, a defect placed at $x=0$ can be introduced in the generating functions of conserved quantities in the following form,
\br
 {\cal I}_j = \int_{-\infty}^0 d x\,\left[V^{(1)}_{jj} + \sum_{i\neq j} V^{(1)}_{ji} \G_{ij}[\P^{(1)}] \] + \int_0^{\infty}d x\,\left[V^{(2)}_{jj} + \sum_{i\neq j} V^{(2)}_{ji} \G_{ij}[\P^{(2)}]  \], \label{eq4.39}
\er
where $V_{ij}^{(p)}$ with $p=1,2$ are the components of the $x$-part of the Lax connections (\ref{problema2}) describing each associated linear problem in the regions $x<0$, and $x>0$ respectively; and $\G_{ij}[\P^{(p)}] =\Psi^{(p)}_{i}(\Psi_j^{(p)})^{-1}$ are their corresponding set of auxiliary functions derived in the last section. After taking the time-derivative of (\ref{eq4.39}), we find that
the modified quantities ${\cal I}_j+D_j$ are conserved, where
\br
 \qquad D_j = \ln\left[ K_{jj} + \sum\limits_{i\neq j} K_{ji}\G_{ij}[\P^{(1)}] \right]\Bigg|_{x=0},\label{defcont}
\er
gives the defect contributions to the $j$-th generating function of conserved quantities, and its precise form depends on the components of the defect matrix. From the above formula (\ref{defcont}) we will derive two different sets of defect contributions by considering the expansion of the auxiliary functions in positive and negative powers of $\l$ respectively. 

Firstly, from the explicit form of the defect matrix (\ref{K caso bf}), and the coefficients correspon\-ding to the expansions in positives powers of $\l$ (\ref{eqn4.10})--(\ref{eqn4.13}) and (\ref{eqn4.24})--(\ref{eqn4.26}), we get
\begin{eqnarray}
D_{1}^{(0)}&=&(\ph_{2}-\ph_{1}), \qquad D_{2}^{(0)}=(\ph_{1}-\ph_{2}),\\
D_{1}^{(1/2)}&=&\frac{1}{2}\pp_{-}\ph_{1}-\frac{\omega^2}{2}e^{\ph_{1}-\ph_{2}}+i\omega\, e^{\frac{\ph_{1}-\ph_{2}}{2}}f_1 \p_{1},\\
D_{2}^{(1/2)}&=&-\frac{1}{2}\pp_{-}\ph_{1}-\frac{\omega^2}{2}\,e^{-(\ph_{1}-\ph_{2})}+i\omega\, e^{-\frac{(\ph_{1}-\ph_{2})}{2}}f_1 \p_{1}.
\end{eqnarray}
Analogously, from the coefficients for the negative powers of $\l$ (\ref{eqn4.17})--(\ref{eqn4.19}), and (\ref{eqn4.30})--(\ref{eqn4.33}), we find
\begin{eqnarray}
\hat{D}_{1}^{(-1/2)}&=&\left[-\frac{2}{\omega^2}e^{\ph_{1}+\ph_{2}}+\frac{2}{i\o}e^{\frac{\ph_{1}+\ph_{2}}{2}}f_1\bp_{1}\right],\\
\hat{D}_{2}^{(-1/2)}&=&\left[-\frac{2}{\omega^2}e^{-(\ph_{1}+\ph_{2})}+\frac{2}{i\o}e^{-\frac{(\ph_{1}+\ph_{2})}{2}}f_{1}\bp_{1}\right].
\end{eqnarray} 
Now, let us define the following quantities, 
\begin{eqnarray}
\mathbb{D}^{(+1/2)}=D_{1}^{(1/2)} + D_{2}^{(1/2)}. \qquad \mathbb{D}^{(-1/2)}=\hat{D}_{1}^{(-1/2)}+\hat{D}_{2}^{(-1/2)}. \label{eqn4.47}
\end{eqnarray}
Then, the defect contributions to the modified energy and momentum are recovered by adding and subtracting the expression in (\ref{eqn4.47}) as follows,
\be
E_{D}&=&\hat{\mathbb{D}}^{(-1/2)}+ \mathbb{D}^{(+1/2)}=-\o^{2}\cosh(\ph_{1}-\ph_{2})-
\frac{4}{\o^{2}}\cosh(\ph_{1}+\ph_{2})\non\\&&+
2i\o\cosh\left(\frac{\ph_{1}-\ph_{2}}{2}\right)f_{1}\p_{1}-
\frac{4i}{\o}\cosh\left(\frac{\ph_{1}+\ph_{2}}{2}\right)f_{1}\bp_{1},\\
P_{D}&=&\hat{\mathbb{D}}^{(-1/2)}-\mathbb{D}^{(+1/2)}=\o^{2}\cosh(\ph_{1}-\ph_{2})-
\frac{4}{\o^{2}}\cosh(\ph_{1}+\ph_{2})\non\\&&-
2i\o\cosh\left(\frac{\ph_{1}-\ph_{2}}{2}\right)f_{1}\p_{1}-
\frac{4i}{\o}\cosh\left(\frac{\ph_{1}+\ph_{2}}{2}\right)f_{1}\bp_{1}.
\ee
The expressions derived from the Lagrangian formalism in (\ref{eqn2.15}) and (\ref{eqn2.17}) can be reached by noting that the auxiliary field $f_1$ introduced in (\ref{eqn3.19}) satisfy the following relations,
\begin{eqnarray}
\psi_1\psi_2 =-2i\omega{\cosh\Big(\frac{\phi_-}{2}\Big)}f_1\psi_2, \qquad \bpsi_1\bpsi_2
=-\frac{4i}{\omega}\cosh\Big(\frac{\phi_-}{2}\Big)f_1\bpsi_2.
\end{eqnarray}
Then, we finally get
\be
P_{D}&=& \,\,\,\left[\o^{2}\cosh(\ph_{1}-\ph_{2})-
\frac{4}{\o^{2}}\cosh(\ph_{1}+\ph_{2})\right]+(\bp_{2}\bp_{1}+\p_{1}\p_{2})\non\\&&\!\!\!\, +\left[
2i\o\cosh\left(\frac{\ph_{1}-\ph_{2}}{2}\right)(\p_{1}-\p_{2})+
\frac{4i}{\o}\cosh\left(\frac{\ph_{1}+\ph_{2}}{2}\right)(\bp_{1}+\bp_{2})\right]f_1, \\
E_{D}&=&\!\!-\left[\o^{2}\cosh(\ph_{1}-\ph_{2})+
\frac{4}{\o^{2}}\cosh(\ph_{1}+\ph_{2})\right]+(\bp_{2}\bp_{1}-\p_{1}\p_{2})\non\\&&\!\!+
\left[\frac{4i}{\o}\cosh\left(\frac{\ph_{1}+\ph_{2}}{2}\right)(\bp_{1}+\bp_{2}) -2i\o\cosh\left(\frac{\ph_{1}-\ph_{2}}{2}\right)(\p_{1}-\p_{2})\right]f_1.
\ee


\section{Further solutions for the defect matrix}

Notice that eqns. (\ref{gauge}) and (\ref{K})  led to solutions partially of type-II in the sense that  the bosonic part is of type-I (i.e., supersymmetric extension of type-I with an auxiliary fermionic field).  The question we raise in whether (\ref{K}) generates  more general solutions.  In order to face this problem let us consider the pure bosonic limit when the fermions are set to zero.  We shall see that in this case, we obtain solutions of type-II.  In such limit the gauge potentials  are written as
\br
\mathcal{A}_{+}&=&\left(\begin{array}{ccc}\lambda^{1/2}-\pp_{+}\phi&-1&0\\-\lambda&\lambda^{1/2}+\pp_{+}\phi&0\\0&0&2\lambda^{1/2}\end{array}\right),\non\\ 
\mathcal{A}_{-}&=&\left(\begin{array}{ccc}\lambda^{-1/2}&-\lambda^{-1}e^{2\phi}&0\\- e^{-2\phi}&\lambda^{-1/2}&0\\0&0&2\lambda^{-1/2}\end{array}\right)\ .\ 
\er
From (\ref{dm5}) and (\ref{dm12})  we obtain directly that
\br
\gamma_{12}&=& 0,\quad\beta_{21}=0,\\
\gamma_{22}&=&\gamma_{11},\quad\beta_{22}=\beta_{11}e^{2\phi_-}\label{r1} \ .
\er
and since $\bar{\psi}=0$ we find from (\ref{dm6})-(\ref{dm11}) 
\br
	\alpha_{22}&=&\alpha_{11},\quad\gamma_{23}=-\alpha_{13},\quad\gamma_{31}=-\alpha_{32}\label{r2}\\
	\alpha_{12}&=& 0,\quad\gamma_{13}=0\quad\gamma_{32}=0\ .
\er
Similarly with $\psi=0$ we have from (\ref{dm13})-(\ref{dm18}) 
\br
	\beta_{13}&=&-e^{(\phi_+-\phi_-)}\alpha_{23},\quad\beta_{32}=-e^{(\phi_+ +\phi_-)}\alpha_{31},\quad\alpha_{22}=\alpha_{11}e^{2\phi_-}\label{r3}\\
	\alpha_{21}&=& 0,\quad\beta_{23}=0,\quad\beta_{31}=0
\er
Considering the equations involving $\b_{11}$(\ref{dm39}), $\a_{23}$(\ref{dm21}), $\a_{31}$(\ref{dm23}), i.e,
\br
\pa_+\b_{11} = -\b_{11} \pa_+\phi_-, \quad \pp_{+}\a_{23}=-\frac{\a_{23}}{2}\pp_{+}({\phi_+-\phi_-}),\quad\pp_{+}\a_{31}=-\frac{\a_{31}}{2}\pp_{+}({\phi_++\phi_-}).
\er
and with the results above (\ref{r1}), (\ref{r2}) and (\ref{r3}) we find the simple solutions
\be
\b_{11}&=& b_{11}e^{-\phi_-},\quad\b_{22}= b_{11}e^{\phi_-},\quad\a_{23}=a_{23}\,e^{-\frac{(\phi_+-\phi_-)}{2}}\non\\
\a_{31}&=& a_{31}e^{-\frac{(\phi_++\phi_-)}{2}},\quad\b_{13}=-a_{23}\,e^{\frac{(\phi_+-\phi_-)}{2}},\quad
\b_{32}=-a_{31}\,e^{\frac{(\phi_++\phi_-)}{2}}.
\ee
Whereas $\a_{11}=\a_{22}$ and $\g_{11}=\g_{22}$, then the equations for these elements (\ref{dm19}), (\ref{dm22}), (\ref{dm45}), (\ref{dm50}), (\ref{dm51}), has the solutions $\a_{11}=\a_{22}=0$ and $\g_{11}=\g_{22}=c_{11}$, respectively. From equations (\ref{dm33}) and (\ref{dm50}) involving $\b_{11}$ and $\g_{11}$, we get
\begin{eqnarray}
\pp_{-}\phi_- &=&\frac{1}{b_{11}}(\b_{12}e^{-\phi_+} -\g_{21}e^{\phi_+})\label{b11},\\
\pp_{+}\phi_- &=&\frac{1}{c_{11}}(\g_{21}-\b_{12}).\label{c11}
\end{eqnarray}
Choosing the fermionic constants $(a_{23}, a_{13}, a_{31}, a_{32})$ to vanish, the system of equations (\ref{dm34}), (\ref{dm40}), (\ref{dm46}), (\ref{dm52}) reduces to
\begin{eqnarray}
\pp_{+}\g_{21}&=&-\g_{21}\pp_{+}\phi_+-b_{11}(e^{\phi_-}-e^{-\phi_-}),\label{c21+}\\
\pp_{-}\g_{21}&=&c_{11}e^{-\phi_+}(e^{\phi_+}-e^{-\phi_-}),\label{c21-}\\
\pp_{+}\b_{12}&=&\b_{12}\pp_{+}\phi_+ +b_{11}(e^{\ph_{-}}-e^{-\phi_-}),\label{b12+}\\
\pp_{-}\b_{12}&=&-c_{11}e^{\phi_+}(e^{\phi_-}-e^{-\phi_-})\label{b12-} .
\end{eqnarray}
Introducing the auxiliary field $\Lambda$ such that
\begin{eqnarray*}
	\g_{21}&=&c_{21}e^{\L-\phi_+},
\end{eqnarray*}
equations (\ref{b11})-(\ref{c21-}) become
\begin{eqnarray}
\pp_{+}\phi_-&=&\frac{1}{c_{11}}(c_{21}e^{\Lambda-\phi_+}-\beta_{12}),\label{ph+}\\
\pp_{-}\phi_-&=&-\frac{1}{b_{11}}(c_{21}e^{\Lambda}-e^{-\phi_+}\beta_{12}),\label{ph-}\\
\pp_{+}\Lambda&=&-\frac{b_{11}}{c_{21}}e^{\phi_+ -\Lambda}(e^{\phi_-}-e^{-\phi_-}),\label{l}\\
\pp_{-}(\Lambda-\phi_+)&=&\frac{c_{11}}{c_{21}}e^{-\Lambda}(e^{\phi_-}-e^{-\phi_-} )\label{l+p}.
\end{eqnarray}
A compatible solution for (\ref{ph+})-(\ref{l+p}) and (\ref{b12+}), (\ref{b12-}) is
\be
\beta_{12}&=c_{21}e^{(\phi_+ -\Lambda )}(e^{\phi_-}+e^{-\phi_-}+\eta)\ .
\ee
The above system is identified as  a
type-II Backlund transformation for the bosonic  sinh-Gordon model \cite{Corr09}, \cite{Ale3}.
Therefore the matrix $K$ is given now by
\begin{eqnarray*}
	\cK=\left(\begin{array}{ccc}
		\lambda^{-1/2}b_{11}e^{-\phi_-}+\lambda^{1/2}c_{11} & \lambda^{-1/2}c_{21}e^{\phi_+-\Lambda}(e^{\phi_-}+e^{-\phi_-}+\eta) & 0 \\ 
		\lambda^{1/2}c_{21}e^{\Lambda-\phi_+} & \lambda^{-1/2}b_{11}e^{\phi_-}+\lambda^{1/2}c_{11} & 0 \\ 
		0 & 0 & a_{33}+\lambda^{-1/2}b_{33}+\lambda^{1/2}c_{33}
	\end{array} \right)\! .
\end{eqnarray*}
This result for the pure bosonic case indicates that eqns. (\ref{gauge}) and (\ref{K})  may also generate \mbox{type-II} Backlund transformations for the supersymmetric system, which is a subject under our investigation.

\vskip 1cm
 \noindent
{\bf Acknowledgements} \\
\vskip .1cm \noindent
{ARA  thanks Fapesp. NIS, JFG and AHZ  thank CNPq for partial support.}

\newpage
\appendix
\section{Supersymmetry of the $N=1$ defect sshG model}
\label{appA}

The action for the bulk $N=1$ sshG model,
\br
 S_{bulk} = \int_{-\infty}^{\infty} dt \int_{-\infty}^{\infty}  dx \ {\cal L}_{bulk},
\er
with,
\begin{eqnarray}
 {\mathcal L}_{bulk} &=& \frac{1}{2}(\partial_x \phi)^2 - \frac{1}{2}(\partial_t \phi)^2 + \bar{\psi}(\partial_t -\partial_x)\bar{\psi} +   \psi(\partial_t +\partial_x)\psi + 4\cosh(2\phi) +8\bpsi\psi\cosh\phi, \qquad \mbox{}\label{ssA2}
\end{eqnarray}
has $N=1$ supersymmetry without topological charge \cite{VF77}, \cite{Hruby}. The supersymmetry transformation is given by
\br
 \d_s\phi &=& \vep\,\psi+\bvep\,\bpsi, \label{ssA3}\\
 \d_s\psi &=& -\vep \,\pa_- \phi + 2\bvep\sinh\phi,\\
 \d_s\bpsi &=& \bvep \,\pa_{+}\phi - 2\vep \sinh\phi , \label{ssA5}
\er
where $\vep$ and $\bvep$ are fermionic parameters. Under a general  not-rigid susy transformation, i.e with parameters  $\vep(x,t)$ and $\bvep(x,t)$,   ${\cal L}_{bulk}$ changes by a total derivative 
\br
 \d_s {\cal L}_{bulk} &=& \vep \Big[\pa_t\left(-2\psi\,\pa_-\phi-2\bpsi\,\sinh\phi\) +\pa_x\left(-2\psi\,\pa_-\phi+4\bpsi\,\sinh\phi\) \Big]\non \\
 &&\!\!\! \!\!+\,\bvep \Big[\pa_t\left(2\bpsi\,\pa_+\phi+4\bpsi\sinh\phi\) +\pa_x\left(-2\bpsi\,\pa_+ \phi+4\bpsi\,\sinh\phi\)\Big]\non \\
 && \!\!\! \!\!+\pa_x\Big[\vep \psi (\pa_+\phi + 2\pa_-\phi)+\bvep \bpsi(\pa_-\phi+2\pa_+\phi)-2(\vep\bpsi+\bvep\psi)\sinh\phi\Big]\non \\
 && \!\!\! \!\!+\pa_t\Big[-\vep \psi (\pa_+\phi - 2\pa_-\phi)+\bvep \bpsi(\pa_-\phi-2\pa_+\phi)+2(\vep\bpsi-\bvep\psi)\sinh\phi\Big],
\er
if the conservation laws holds,
\br
\pa_t\left(-2\psi\,\pa_-\phi-2\bpsi\,\sinh\phi\) +\pa_x\left(-2\psi\,\pa_-\phi+4\bpsi\,\sinh\phi\) &=&0, \\
\pa_t\left(2\bpsi\,\pa_+\phi+4\bpsi\sinh\phi\) +\pa_x\left(-2\bpsi\,\pa_+ \phi+4\bpsi\,\sinh\phi\)&=&0.
\er
Then, the associated bulk supercharges $Q_{\vep}$ and $\bar{Q}_{\bvep}$ can be written as integrals of local fermionic densities, as follows
\br
Q_{\vep} =- \int_{-\infty}^{\infty}dx\,\left(2\psi\,\pa_-\phi + 4\bpsi\sinh\phi\),\qquad
 \bar{Q}_{\bvep} =\int_{-\infty}^{\infty}dx\,\left(2\bpsi\,\pa_{+}\phi + 4\psi\sinh\phi\).
\er
From the above expressions can be easily verified that $\{Q_{\vep},\bar{Q}_{\bvep}\}=0$. On the other hand, when considering a rigid (constant parameters) susy transformation, ${\cal L}_{bulk}$ changes as
\br
 \d_s {\cal L}_{bulk} &=& \pa_x\left[\frac{1}{2}(\pa_x\phi)(\vep \psi +\bvep \bpsi) +\frac{1}{2}(\pa_t\phi)(\vep\psi-\bvep\bpsi)+2(\bvep\psi+\vep\bpsi)\sinh\phi\right]\non \\
 &&\!\!\! \!\!+\ \pa_t\left[\frac{1}{2}(\pa_x\phi)(\bvep \bpsi-\vep \psi ) -\frac{1}{2}(\pa_t\phi)(\vep\psi+\bvep\bpsi)+2(\bvep\psi-\vep\bpsi)\sinh\phi\right].
\er
It turns to be that the bulk theory defined by the Lagrangian (\ref{ssA2}) is invariant under susy transformation. However, this is not necessarily true for the defect theory, and therefore we should show that the presence of the defect will not destroy the supersymmetry of the bulk theory. In fact, in the defect theory the total action (left half-line $+$ defect $+$ right half-line),
\br
 S_{total}= \int_{-\infty}^{\infty} dt \left[\int_{-\infty}^{0}  dx \ {\cal L}_1 +\int_{0}^{\infty}  dx \ {\cal L}_2 +{\cal L}_D\right],
\er
with
\begin{eqnarray}
 {\mathcal L}_p &=& \frac{1}{2}(\partial_x \phi_p)^2 - \frac{1}{2}(\partial_t \phi_p)^2 + \bar{\psi}_p(\partial_t -\partial_x)\bar{\psi}_p +   \psi_p(\partial_t +\partial_x)\psi_p  + 4\cosh(2\phi_p) \non \\&&+8\bpsi_p\psi_p\cosh\phi_p,
\end{eqnarray}
under the susy transformation (\ref{ssA3})--(\ref{ssA5}) changes as follows,
\br
 \d_s {S}_{total} &=&\!\! \int_{-\infty}^{\infty} dt \Bigg[\left(\frac{1}{2}(\pa_x\phi_1)(\vep \psi_1 +\bvep \bpsi_1) +\frac{1}{2}(\pa_t\phi_1)(\vep\psi_1-\bvep\bpsi_1)+2(\bvep\psi_1+\vep\bpsi_1)\sinh\phi_1\right)_{x=0}\non\\
 &&\qquad \,\,\, -\left(\frac{1}{2}(\pa_x\phi_2)(\vep \psi_2 +\bvep \bpsi_2) +\frac{1}{2}(\pa_t\phi_2)(\vep\psi_2-\bvep\bpsi_2)+2(\bvep\psi_2+\vep\bpsi_2)\sinh\phi_2\right)_{x=0}\non\\&&\qquad \quad +\,\d {\cal L}_D\Big].\qquad \,\,\,\mbox{}\label{ssA13}
\er
Then, as the right-hand-side of the above equation does not vanish immediately, we should show that the variation of the defect Lagrangian, 
\br
 {\mathcal L}_D &=&\frac{1}{2}(\phi_2\partial_t\phi_1-\phi_1\partial_t\phi_2) -\psi_1\psi_2 -\bar{\psi}_1\bar{\psi}_2 + 2f_1\pa_t f_1+B_0 +B_1, \label{ssA14}
\er
cancels out the surface terms and exactly restores supersymmetry, where the defect potentials $B_k$, with $k=0,1$ are given by
\br 
B_0 &=&  -\frac{4}{\om^2} \cosh(\phi_1+\phi_2) -\om^2\cosh(\phi_1-\phi_2),\\
 B_1&=&-\frac{4i}{\om}\cosh\Big(\frac{\phi_1+\phi_2}{2}\Big)f_1(\bar{\psi}_1+\bar{\psi}_2) +2i\om\,\cosh\Big(\frac{\phi_1-\phi_2}{2}\Big) f_1(\psi_1 -\psi_2).
\er
It is important to take into account that all the fields appearing in the defect Lagrangian are valued in $x=0$ and only depend on time. Now, let us first derive the corresponding susy transformation for the auxiliary fermionic field $f_1$. By applying the susy variation on (\ref{defeito3}), one of the defect equations involving $f_1$, namely
\br
\psi_{1}+\psi_{2} & = & -2i\om\,\cosh\Big(\frac{\phi_{1}-\phi_{2}}{2}\Big)f_{1},
\er
we get from l.h.s.,
\br
\d_s (\psi_{1}+\psi_{2})&=& -\vep \pa_-(\phi_1+\phi_2) +2\bvep (\sinh\phi_1 +\sinh\phi_2)\non \\
&=&  -\vep \pa_-(\phi_1+\phi_2) +4\bvep \sinh\Big(\frac{\phi_1+\phi_2}{2}\Big)\cosh\Big(\frac{\phi_1-\phi_2}{2}\Big), 
\er
and from the r.h.s. we find,
\br
\d_s (\psi_{1}+\psi_{2}) =-i\vep\,\omega \sinh\Big(\frac{\phi_1-\phi_2}{2}\Big)(\psi_1-\psi_2)f_1-2i\omega \cosh\Big(\frac{\phi_1-\phi_2}{2}\Big)(\d_sf_1),
\er
where we have used eq.(\ref{defeito4}). Now, comparing the above results and using eq. (\ref{eqn3.33}) it can be checked that the susy transformation of the fermionic field $f_1$ is given by
\br 
\d_s f_1 &=& -i\omega \,\vep\sinh\(\frac{\phi_1-\phi_2}{2}\)  + \frac{2i \bvep}{\omega}\sinh\(\frac{\phi_1+\phi_2}{2}\).
\er
Now, we find that the supersymmetry variations of the bosonic terms of the defect Lagrangian (\ref{ssA14}) are,
\br 
 \d_s\(\frac{1}{2}(\phi_2\partial_t\phi_1-\phi_1\partial_t\phi_2)\)&=& (\vep\psi_2+\bvep \bpsi_2) (\pa_t\phi_1) - (\vep\psi_1+\bvep \bpsi_1)(\pa_t\phi_2), \non \\
 &=& \vep\left[-\psi_1\pa_t\phi_1 +\psi_2\pa_t\phi_2 -2i\omega \cosh\Big(\frac{\phi_1-\phi_2}{2}\Big)f_1 \pa_t(\phi_1-\phi_2)\]\non \\
 &&\!\!\!\!\! + \,\bvep\left[\bpsi_1\pa_t\phi_1 - \bpsi_2\pa_t\phi_2 +\frac{4i}{\omega} \cosh\Big(\frac{\phi_1+\phi_2}{2}\Big)f_1 \pa_t(\phi_1+\phi_2) \right]\!\!, \qquad \,\,\,\mbox{}
\er
and
\br
\d_s B_0 &=&\vep\left[ -\frac{4}{\om^2} \sinh(\phi_1+\phi_2) (\psi_1+\psi_2) -\om^2\sinh(\phi_1-\phi_2)(\psi_1-\psi_2)\right],\\
&& \!\!\!\!\! + \,\bvep\left[-\frac{4}{\om^2} \sinh(\phi_1+\phi_2) (\bpsi_1+\bpsi_2)-\om^2\sinh(\phi_1-\phi_2)(\bpsi_1-\bpsi_2)\].
\er
For the terms involving only fermionic fields, we obtain
\br
 \d_s (-\psi_1\psi_2 -\bar{\psi}_1\bar{\psi}_2) &=& -(\d_s \psi_1)\psi_2+(\d_s \psi_2)\psi_1-(\d_s \bpsi_1)\bpsi_2+(\d_s \bpsi_2)\psi_1, \non \\
 &=& \vep\Big[-\psi_1\pa_-\phi_1+\psi_2\pa_-\phi_2 + 2(\bpsi_2\sinh\phi_1 -\bpsi_1\sinh\phi_2)\non\\ &&\quad -\,2i\omega\cosh\Big(\frac{\phi_1-\phi_2}{2}\Big)f_1 \pa_-(\phi_1-\phi_2)\Big]+ \bvep\Big[-\bpsi_1\pa_+\phi_1+\bpsi_2\pa_+\phi_2 \non \\&& \!\!\!\!\! - \, 2(\psi_2\sinh\phi_1 -\psi_1\sinh\phi_2)-\frac{4i}{\omega}\cosh\Big(\frac{\phi_1+\phi_2}{2}\Big)f_1 \pa_-(\phi_1+\phi_2)\Big]\! , \qquad \,\,\,\mbox{}
\er
and
\br
\d_s(2f_1\pa_tf_1)\!\!\!&=&\!\!\!\left[2i\omega \vep\cosh\Big(\frac{\phi_1-\phi_2}{2}\Big) \pa_t(\phi_1-\phi_2)-\frac{4i \bvep}{\omega}\cosh\Big(\frac{\phi_1+\phi_2}{2}\Big) \pa_t(\phi_1+\phi_2)\right]\!f_1.\qquad\mbox{}
\er
In addition, for the defect potential $B_1$ we get,
\br
 \d_s B_1 &=&-\frac{4i}{\om}\cosh\Big(\frac{\phi_1+\phi_2}{2}\Big)\d_s\left[f_1(\bar{\psi}_1+\bar{\psi}_2)\right] +2i\om\,\cosh\Big(\frac{\phi_1-\phi_2}{2}\Big) \d_s\left[ f_1(\psi_1 -\psi_2)\right]\non\\
 &=& \vep\Big[2(\psi_1\pa_-\phi_2-\psi_2\pa_-\phi_1) +4(\bpsi_1\sinh\phi_2-\bpsi_2\sinh\phi_1)\Big]\non \\
 && \!\!\!\!\! + \,\bvep\Big[2(\bpsi_2\pa_+\phi_1-\psi_1\pa_+\phi_2) + 4(\psi_2\sinh\phi_1-\psi_1\sinh\phi_2)\Big].
\er

Now, by putting together all the variations obtained we can find after some algebra that the susy transformation of the defect Lagrangian is exactly given by,
\br
\d_s{\cal L}_D &=& \vep\Big[\psi_1\pa_+\phi_1-\psi_2\pa_+\phi_2 +2(\bpsi_1\sinh\phi_1-\bpsi_2\sinh\phi_2) \Big]_{x=0} \non \\
&& \!\!\!\!\! + \,\bvep\Big[\bpsi_1\pa_-\phi_1-\bpsi_2\pa_-\phi_2 +2(\psi_1\sinh\phi_1-\psi_2\sinh\phi_2)\Big]_{x=0},
\er
which cancels out exactly the surface terms in (\ref{ssA13}). Then, we have shown that the defect sshG model has a well-defined $N=1$ supersymmetry. This implies that there must be modified  supercharges which are preserved. Let us compute the defect contribution for $Q_{\vep} $. After the introduction of the defect at $x=0$, we get
\br
 Q_{\vep} &=&- \int_{-\infty}^{0}dx\,\left(2\psi_1\,\pa_-\phi_1 + 4\bpsi_1\sinh\phi_1\) -\int_{0}^{\infty}dx\,\left(2\psi_2\,\pa_-\phi_2 + 4\bpsi_2\sinh\phi_2\).\qquad \mbox{}
\er
Now, by taking the time-derivative respectively, we get
\br
\frac{d Q_{\vep}}{dt} &=& \Big[\psi_1(\pa_x\phi_1-\pa_t\phi_1) - 4\bpsi_2\sinh\phi_1-\psi_2(\pa_x\phi_2-\pa_t\phi_2)+ 4\bpsi_2\sinh\phi_2\Big]_{x=0},\non\\
&=& \Big[-(\psi_1+\psi_2)\pa_t(\phi_1-\phi_2)-4(\bpsi_1\sinh\phi_1-\bpsi_2\sinh\phi_2) +\omega^2 (\psi_1-\psi_2)\sinh(\phi_1-\phi_2)\non\\&& \,\,\,+\frac{4}{\omega^2}(\psi_1+\psi_2)\sinh(\phi_1+\phi_2)\Big]_{x=0}\non\\
&=&\Big[-2(\sinh\phi_1+\sinh\phi_2)(\bpsi_1-\bpsi_2) +\frac{4}{\omega^2}(\psi_1+\psi_2)\sinh(\phi_1+\phi_2)\Big]_{x=0}\non \\
&& \!\!\! +\, \frac{d}{dt}\Big[4i\omega\sinh\Big(\frac{\phi_1-\phi_2}{2}\Big)\,f_1 \Big]_{x=0}. 
\er
The first term in the above result vanishes after using the defect conditions (\ref{defeito3}) and  (\ref{defeito4}). Then we find that the modified conserved supercharges can be written in the form ${\cal Q}= Q_\vep + Q_{\mbox{\tiny D}} $, with the defect contribution given by
 \br
 Q_{\mbox{\tiny D}} &=& -4i\omega\left[\sinh\Big(\frac{\phi_1-\phi_2}{2}\Big)\,f_1 \right]_{x=0}.
\er
Analogously, for the supercharge $\bar{Q}_{\bvep}$,
\br
 \bar{Q}_{\bvep} &=&\int_{-\infty}^{0}dx\,\left(2\bpsi_1\,\pa_{+}\phi_1 + 4\psi_1\sinh\phi_1\)+\int_{0}^{\infty}dx\,\left(2\bpsi_2\,\pa_{+}\phi_2 + 4\psi_2\sinh\phi_2\), \mbox{}
\er
the corresponding modified conserved supercharge is $ \bar{\cal Q} = \bar{Q}_{\bvep} + \bar{Q}_{\mbox{\tiny D}}$, where the defect contribution is given by
\br
\bar{Q}_{\mbox{\tiny D}}= \frac{8i}{\om} \left[\sinh\Big(\frac{\phi_1+\phi_2}{2}\Big)\,f_1 \right]_{x=0}.
\er

\newpage
\section{Calculation of the defect matrix}
\label{appB}
The defect matrix ${\cal K}$ is directly derived by solving the differential equations,
\br
\pa_{\pm}\cK=\cK\mathcal{A}^{(1)}_{\pm}-\mathcal{A}^{(2)}_{\pm}\cK,
\er
with the Lax connections are given 
\br
\mathcal{A}_{+} &=&\left(\begin{array}{cc|c}\lambda^{1/2}-\pa_{+}\phi&-1&\bar{\psi}\\[0.2cm] -\l&\l^{1/2}+\pa_{+}\phi&\l^{1/2}\bpsi \tabularnewline\hline \mbox{}&\mbox{}&\mbox{}\\[-0.3cm]
 \l^{1/2}\bpsi &\bpsi &2\l^{1/2}\end{array}\right), \\[0.2cm]
\mathcal{A}_{-}&=&\left(\begin{array}{cc|c}\lambda^{-1/2} &-\lambda^{-1}e^{2\phi}&-\lambda^{-1/2}\psi \,e^{\phi}\\[0.2cm]- e^{-2\phi}&\lambda^{-1/2}&- \psi e^{-\phi} \tabularnewline\hline \mbox{}&\mbox{}&\mbox{}\\[-0.3cm]
\psi e^{-\phi}&\lambda^{-1/2}\psi \,e^{\phi}&2\lambda^{-1/2}\end{array}\right).
\er
To find a solution for the defect matrix ${\cK}$, we propose the following $\l$-expansion,
\begin{eqnarray}\label{K1}
\cK_{ij}=\a_{ij}+\l^{-1/2}\b_{ij}+\l^{1/2}\g_{ij}.
\end{eqnarray}
Now, considering term by term we find a set of constraints coming from the $\l^{\pm 3/2}$ and $\l^{\pm 1}$ terms, which we will present explicitly as follows:\\\\
\underline{$\l^{+3/2}$- terms :}
\begin{eqnarray}
\g_{12}=\g_{13}=\g_{32}=0, \qquad \quad \g_{11}=\g_{22}. \label{dm5}
\end{eqnarray}
\underline{$\l^{+1}$- terms :} 
\begin{eqnarray}
\a_{12} &=& \g_{13}\,\bpsi_1, \label{dm6}\\
\g_{13} &=& -\g_{12}\,\bpsi_1,\label{dm7}\\
\g_{32} &=&-\g_{12}\,\bpsi_2,\label{dm8}\\
 \a_{11}-\a_{22}&=&\bp_2\,\g_{31}-\g_{23}\,\bp_1,\label{dm9}\\
\a_{13}+\g_{23}&=&\bp_2\,\g_{33}-\g_{11}\,\bp_1,\label{dm10}\\ 
\g_{31}+\a_{32}&=&\bp_1\,\g_{33}-\g_{11}\,\bp_2.\label{dm11}
\er
\underline{$\l^{-3/2}$- terms :} 
\begin{eqnarray}
\b_{21}=\b_{23}= \b_{31}=0, \qquad \quad \b_{22}=\b_{11}e^{2\phi_-}.\label{dm12}
\end{eqnarray}
\underline{$\l^{-1}$- terms :} 
\br
\a_{21} &=& -\psi_2\,\b_{31}\,e^{-\frac{(\phi_+-\phi_-)}{2}},\label{dm13}\\
\b_{23} &=&\b_{21}\,\psi_1\,e^{\frac{(\phi_+ +\phi_-)}{2}},\label{dm14}\\ 
\b_{31} &=&  -\psi_2\,\b_{21}\,e^{\frac{(\phi_+-\phi_-)}{2}},\label{dm15}
\er
\br
\a_{11}\,e^{\phi_-}-\a_{22}\,e^{-\phi_-}&=&e^{-\frac{\phi_+}{2}}\big(\b_{13}\,\p_1\,e^{\frac{\phi_-}{2}}+e^{-\frac{\phi_-}{2}}\,\p_2\,\b_{32}\big),\label{dm16}\\
\a_{31}\,e^{{(\phi_+ +\phi_-)}}+\b_{32}&=& e^{\frac{\phi_+ }{2}}\big(\b_{33}\,\p_1\,e^{\frac{\phi_-}{2}}-\b_{22}\,\p_2\,e^{-\frac{\phi_-}{2}}\big),\label{dm17}\\
\a_{23}\,e^{(\phi_+-\phi_-)}+ \b_{13}&=&e^{\frac{\phi_+}{2}}(\b_{11}\,\p_1e^{\frac{\phi_-}{2}}-\b_{33}\,\p_2 e^{-\frac{\phi_-}{2}}).\label{dm18}
\er
We have denoted  $\phi_{\pm} =\phi_1\pm\phi_2$ for convenience. From eqs (\ref{dm6}), (\ref{dm7}) we get that $\a_{12}=0$, and from (\ref{dm13}) and (\ref{dm14}) that $\a_{21}=0$. 
Now, we obtain a set of differential equations from the $\l^{0}$ and $\l^{\pm 1/2}$ terms. Here we summarize them:\\\\
\underline{$\l^{0}$- terms :} 
\br
\pp_{+}\a_{11}&=&-\a_{11}\,\pp_{+}\phi_- +\b_{13}\,\bp_1-\bp_2\,\a_{31},\label{dm19}\\[0.1cm]  
\pp_{+}\a_{13}&=&\frac{\a_{13}}{2}\,\pp_{+}({\phi_+-\phi_-})+\b_{13}+\a_{23}-\bp_2\,\a_{33}+(\a_{11}+\b_{12})\bp_1,\label{dm20}\\
\pp_{+}\a_{23}&=&-\frac{\a_{23}}{2}\pp_{+}({\phi_+-\phi_-})+\b_{22}\,\bp_1-\bp_2\,\b_{33},\label{dm21}\\
\pp_{+}\a_{22}&=&\a_{22}\,\pp_{+}\phi_- +\a_{23}\,\bp_1-\bp_2\,\b_{32},\label{dm22}\\
\pp_{+}\a_{31}&=&-\frac{\a_{31}}{2}\pp_{+}({\phi_++\phi_-})+\b_{33}\,\bp_1-\bp_2\,\b_{11},\label{dm23}\\
\pp_{+}\a_{32}&=&\frac{\a_{32}}{2}\pp_{+}({\phi_++\phi_-})-\a_{31}-\b_{32}+\a_{33}\,\bp_1-\bp_2(\b_{12}+\a_{22}),\label{dm24}\\
\pp_{+}\a_{33}&=&(\b_{32}+\a_{31})\bp_1-\bp_2(\a_{23}+\b_{13}),\label{dm25}\\[0.1cm]
\pp_{-}\a_{11}&=&e^{-\frac{\phi_-}{2}}(\a_{13}\,\p_1e^{-\frac{\phi_+}{2}}+e^{\frac{\phi_+}{2}}\p_2\,\g_{31}),\label{dm26}\\[0.1cm]
\pp_{-}\a_{13}&=&e^{\frac{\phi_+}{2}}(e^{-\frac{\phi_-}{2}}\,\p_2\,\g_{33}-\g_{11}\,\p_1\,e^{\frac{\phi_-}{2}}), \label{dm27}\\[0.1cm]
\pp_{-}\a_{22}&=&e^{\frac{\phi_-}{2}}(e^{-\frac{\phi_+}{2}}\,\p_2\,\a_{32}+\g_{23}\,\p_1\,e^{\frac{\phi_+}{2}}),\label{dm28}\\[.1cm]
\pp_{-}\a_{23}&=&\g_{23}+\a_{13}\,e^{-(\phi_+-\phi_-)}+e^{-\frac{(\phi_+-\phi_-)}{2}}\p_2\,\a_{33}-(\g_{21}\,e^{\frac{(\phi_++\phi_-)}{2}}+\a_{22}\,e^{-\frac{(\phi_++\phi_-)}{2}})\,\p_1,\qquad \,\,\,\mbox{}\label{dm29}\\[.1cm]
\pp_{-}\a_{31}&=&-\g_{31}-\a_{32}e^{-(\phi_++\phi_-)}+\a_{33}\p_1\,e^{-\frac{(\phi_++\phi_-)}{2}}-\p_2(e^{-\frac{(\phi_+-\phi_-)}{2}}\a_{11}+e^{\frac{(\phi_+-\phi_-)}{2}}\g_{21}),\qquad\mbox{}\label{dm30}\\[.1cm]
\pp_{-}\a_{32}&=&e^{\frac{\phi_+}{2}}(\g_{33}\,\p_1\,e^{\frac{\phi_-}{2}}-e^{-\frac{\phi_-}{2}}\p_2\,\g_{11}),\label{dm31}\\[0.1cm]
\pp_{-}\a_{33}&=&-(\g_{31}e^{\frac{(\phi_++\phi_-)}{2}}+\a_{32}\,e^{-\frac{(\phi_++\phi_-)}{2}})\p_1-\p_2(e^{-\frac{(\phi_+-\phi_-)}{2}}\a_{13}+e^{\frac{(\phi_+-\phi_-)}{2}}\g_{23}).\label{dm32}
\er
\underline{$\l^{-1/2}$- terms :} 
\br
\pp_{-}\b_{11}&=&e^{-\phi_-}(\g_{21}\,e^{\phi_+}-\b_{12}\,e^{-\phi_+})+e^{-\frac{\phi_-}{2}}(\b_{13}\,\p_1\,e^{-\frac{\phi_+}{2}}+e^{\frac{\phi_+}{2}}\,\p_2\,\a_{31}), \label{dm33}\\[.1cm]
\pp_{-}\b_{12}&=&\g_{11}\,e^{\phi_+}(e^{-\phi_-}-e^{\phi_-})+e^{\frac{\phi_+}{2}}(\a_{13\,}\p_1\,e^{\frac{\phi_-}{2}}+e^{-\frac{\phi_-}{2}}\,\p_2\,\a_{32}),\label{dm34}\\[.1cm]
\pp_{-}\b_{13}&=&\g_{23}\,e^{(\phi_+-\phi_-)}+\a_{13}+e^{\frac{(\phi_+-\phi_-)}{2}}\p_2\,\a_{33}-(\a_{11}\,e^{\frac{(\phi_++\phi_-)}{2}}+\b_{12}\,e^{-\frac{(\phi_++\phi_-)}{2}})\,\p_1,\qquad \mbox{}\label{dm35}\\[.1cm]
\pp_{-}\b_{22}&=&e^{\phi_-}(\b_{12}\,e^{-\phi_+}-\g_{21}\,e^{\phi_+})+e^{\frac{\phi_-}{2}}(\a_{23}\,\p_1\,e^{\frac{\phi_+}{2}}+e^{-\frac{\phi_+}{2}}\,\p_2\,\b_{32}),\label{dm36}\\[.1cm]
\pp_{-}\b_{32}&=&-\a_{32}-\g_{31}\,e^{(\phi_++\phi_-)}+\a_{33}\,\p_1\,e^{\frac{(\phi_++\phi_-)}{2}}-\p_2(e^{\frac{(\phi_+-\phi_-)}{2}}\a_{22}+e^{-\frac{(\phi_+-\phi_-)}{2}}\b_{12}),\qquad \mbox{}\label{dm37}\\[0.1cm]
\pp_{-}\b_{33}&=&-(\b_{32}e^{-\frac{(\phi_++\phi_-)}{2}}+\a_{31}e^{\frac{(\phi_++\phi_-)}{2}})\p_1-\p_2(e^{-\frac{(\phi_+-\phi_-)}{2}}\b_{13}+e^{\frac{(\phi_+-\phi_-)}{2}}\a_{23}),\qquad \mbox{}\label{dm38}\\[0.1cm]
\pp_{+}\b_{11}&=&-\b_{11}\,\pp_{+}\phi_-,\label{dm39}\\
\pp_{+}\b_{12}&=&\b_{12}\,\pp_{+}\phi_+ +\b_{22}-\b_{11}-\bp_2\,\b_{32}+\b_{13}\,\bp_1,\label{dm40}\\[0.1cm]
\pp_{+}\b_{13}&=&\frac{\b_{13}}{2}\,\pp_{+}({\phi_+-\phi_-})-\bp_2\b_{33}+\b_{11}\bp_1,\label{dm41}\\%
\pp_{+}\b_{22}&=&\b_{22}\,\pp_{+}\phi_-, \label{dm42}\\[0.1cm]
\pp_{+}\b_{32}&=&\frac{\b_{32}}{2}\pp_{+}({\phi_++\phi_-})-\bp_2\,\b_{22}+\b_{33}\,\bp_1,\label{dm43}\\
\pp_{+}\b_{33}&=&0.\label{dm44}
\er
\underline{$\l^{+1/2}$- terms :} 
\br
\pp_{-}\g_{11}&=&0, \label{dm45}\qquad\\
\pp_{-}\g_{21}&=&\g_{11}\,e^{-\phi_+}(e^{\phi_-}-e^{-\phi_-})+e^{-\frac{\phi_+}{2}}(\g_{23}\,\p_1\,e^{-\frac{\phi_-}{2}}+e^{\frac{\phi_-}{2}}\,\p_2\,\g_{31}),\label{dm46}\\[0.1cm]
\pp_{-}\g_{23}&=&e^{-\frac{\phi_+}{2}}(e^{\frac{\phi_-}{2}}\,\p_2\,\g_{33}-\g_{11}\,\p_1\,e^{-\frac{\phi_-}{2}}),\label{dm47}\\[.1cm]
\pp_{-}\g_{31}&=&e^{-\frac{\phi_+}{2}}(\g_{33}\,\p_1\,e^{-\frac{\phi_-}{2}}-e^{\frac{\phi_-}{2}}\p_2\,\g_{11}), \label{dm48}\\[.1cm]
\pp_{-}\g_{33}&=&0,\qquad \label{dm49}\\[0.1cm]
\pp_{+}\g_{11}&=&-\g_{11}\,\pp_{+}\phi_- +\g_{21}-\b_{12}+\a_{13}\,\bp_1-\bp_2\,\g_{31},\label{dm50}\\[0.1cm]
\pp_{+}\g_{22}&=&\g_{22}\,\pp_{+}\phi_- -\g_{21}+\b_{12}+\g_{23}\,\bp_1-\bp_2\,\a_{32},\label{dm51}\\[0.1cm]
\pp_{+}\g_{21}&=&-\g_{21}\,\pp_{+}\phi_+ +\b_{11}-\b_{22}+\a_{23}\,\bp_1-\bp_2\,\a_{31},\label{dm52}\\[0.1cm]
\pp_{+}\g_{23}&=&-\frac{\g_{23}}{2}\pp_{+}({\phi_+-\phi_-})+\b_{13}+\a_{23}-\bp_2\,\a_{33}+(\a_{22}+\g_{21})\bp_1,\label{dm53}\\
\pp_{+}\g_{31}&=&-\frac{\g_{31}}{2}\pp_{+}({\phi_++\phi_-})-\b_{32}-\a_{31}+\a_{33}\,\bp_1-\bp_2(\a_{11}+\g_{21}),\label{dm54}\\
\pp_{+}\g_{33}&=&(\a_{32}+\g_{31})\bp_1-\bp_2(\a_{13}+\g_{23}).\label{dm55}
\end{eqnarray}

\newpage


\end{document}